# THE MICROWAVE AMPLITUDE AND PHASE SETTING BASED ON EVENT TIMING FOR THE DCLS*

J.F. Zhu[†], H.L. Ding[1][‡], H.K. Li[1], J.W. Han, X.W. Dai, B. Xu[1], L. Shi[1], J.Y. Yang[1], W.Q. Zhang[1]
Institute of Advanced Science Facilities, Shenzhen, China
[1]also at Dalian Institute of Chemical Physics, Chinese Academy of Sciences, Dalian, China

*Abstract*

The primary accelerator of DCLS (Dalian Coherent Light Source) operates at a repetition rate of 20 Hz now, and the beam is divided at the end of the linear accelerator through Kicker to make two 10 Hz beamlines work simultaneously. For the simultaneous emission FEL of two beamlines, the beam energy of the two beamlines is required to be controlled independently, so we need to set the amplitude and phase of each beamline. This paper implements a microwave amplitude and phase setting function based on event timing. We upgraded the EVG/EVR event timing system and LLRF (Low-Level Radiofrequency) system. Two special event codes and a repetition rate division of 10 Hz are added to the event timing system, and we can set the microwave amplitude and phase by judging the event code in LLRF. We ultimately perform the microwave triggering at a repetition rate of 10 Hz for each beamline and validate this function through beam experiments.

## INTRODUCTION TO THE DCLS LINEAR ACCELERATOR

Figure 1 shows the diagram of the DCLS linear accelerator [1]. There is one electron gun (GUN) and seven S-band accelerating tubes (A0 ~ 6). The beam energy at the end of the linear accelerator reaches 300 MeV for modulation and FEL output. As shown in Fig. 2, two 50 MW klystrons (klystron 1 ~ 2) and two 80 MW klystrons (klystron 3 ~ 4) provide high-power microwaves. We deploy four LLRF systems to control microwaves' amplitude and phase. Each LLRF outputs excitation to drive four Solid-State Amplifiers (SSA). Two ICTs (Integrated Current Transformer) realize beam charge monitor, and eight SBPMs (Stripline Beam Position Monitor) are for beam position acquisition. Moreover, three profiles behind bend magnets help us monitor energy and energy spread after accelerator units.

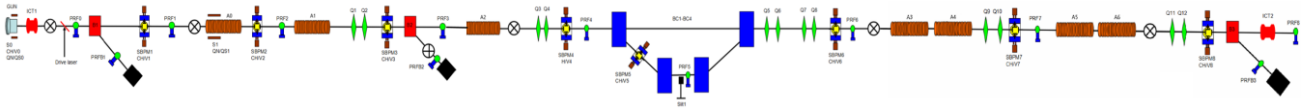

Figure 1: The diagram of the DCLS linear accelerator.

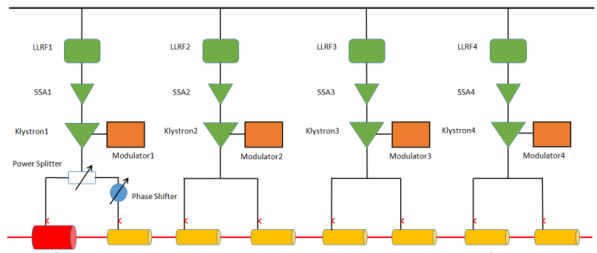

Figure 2: The diagram of the DCLS microwave system.

The timing system is based on the EVG/EVR event timing method [2]. As presented in Fig. 3, the EVG (Event Generator) reference is from the synchronization system. Several EVRs (Event Receivers) receive event codes and provide triggers for other systems. The RMS jitter of the trigger is less than 10 ps. Furthermore, the EVG can be locked by the MPS/PPS protection system. The equipment of the timing system is shown in Fig. 4.

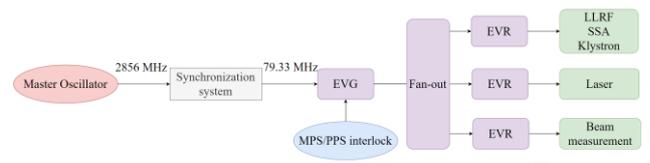

Figure 3: The diagram of the DCLS timing system.

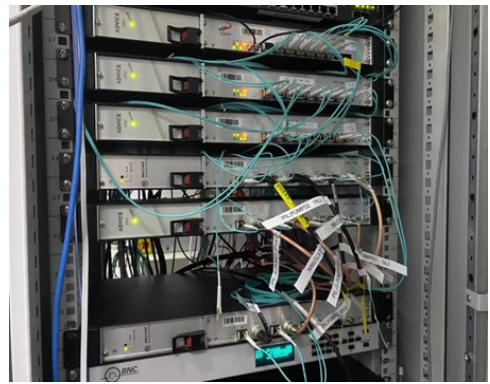

Figure 4: The equipment of the timing system.

___
* Work supported by the National Natural Science Foundation of China (Grant No. 22288201), the Scientific Instrument Developing Project of the Chinese Academy of Sciences (Grant No. GJJSTD20220001), and the Shenzhen Science and Technology Program (Grant No. RCBS20221008093247072).
† zhujinfu@mail.iasf.ac.cn
‡ dinghongli@dicp.ac.cn

## THE TIMING SYSTEM UPGRADE FOR TWO BEAMLINE OPERATIONS

We upgraded the EVG/EVR event timing system and LLRF (Low-Level Radiofrequency) system. As shown in Fig. 5, two special event codes (0xC0 and 0xC1) and a repetition rate division of 10 Hz are added to the event timing system, and we can set the microwave amplitude and phase by judging the event code in LLRF.

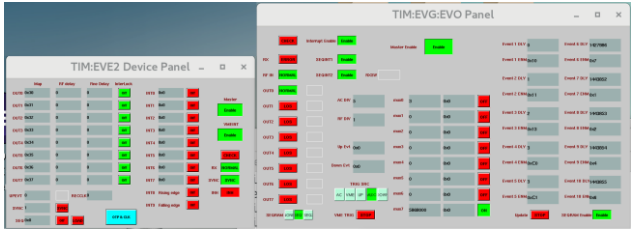

Figure 5: The configuration GUI of the timing system.

In addition, two configurable waveform tables are realized to set the amplitude and phase of microwaves for two beamlines.

## EXPERIMENTS AND RESULTS

The linear accelerator of DCLS operates at a repetition rate of 20 Hz, and the beam is divided at the end of the linear accelerator through Kicker to make two 10 Hz beamlines work simultaneously. We performed the microwave triggering at a repetition rate of 10 Hz for each beamline and validated this function through beam experiments.

Figure 6 depicts the microwave amplitude and phase of two beamlines and electron beams in profile YaG. Specifically, (a) shows the different pulse amplitude and phase setting of Klystron 4 forward for beamlines 1 and 2; (b) shows the electron beams in profile 3 with the exposure time 110 ms when two amplitude microwaves for two beamlines are enabled; (c) shows when the microwave for beamline 1 is disabled; (d) shows when the microwave for beamline 2 is disabled.

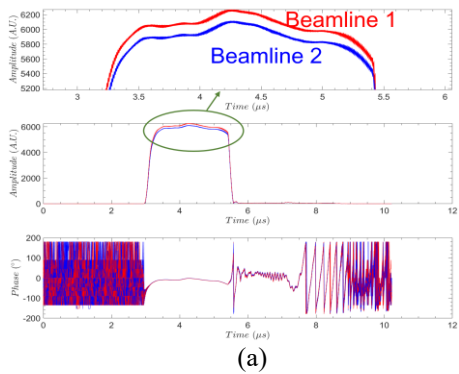

(a)

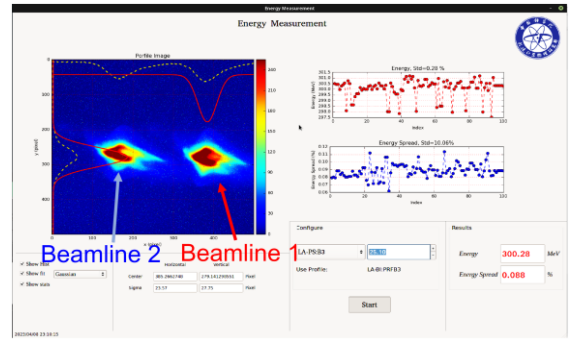

(b)

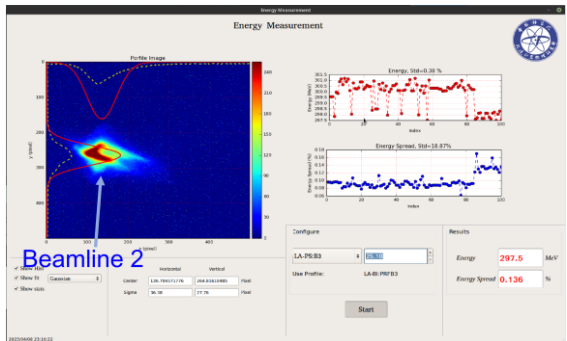

(c)

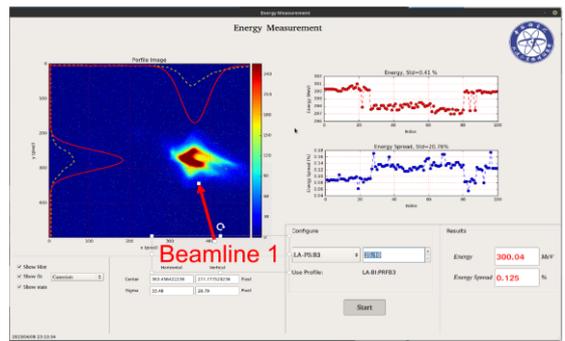

(d)

Figure 6: The microwave amplitude and phase of two beamlines and electron beams shown in profile YaG.

## CONCLUSION

This paper introduces the microwave amplitude and phase setting function based on event timing. The timing and LLRF systems are upgraded, and beam experiments verify the correctness of the function. It can be used for DCLS two-beamline operation.

## ACKNOWLEDGEMENTS

The authors would like to thank the Institute of High Energy Physics, Shanghai Advanced Research Institute, DESY, etc. who collaborated with the DCLS for their support and various discussions over the years.